\documentclass[aps,floatfix,twocolumn,superscriptaddress]{revtex4-2}
\usepackage{epsfig}
\usepackage{color}
\usepackage{amssymb}
\usepackage{amsmath}
\usepackage[ngerman,english]{babel}
\usepackage{multirow}
\usepackage{graphicx}
\usepackage{siunitx}
\newcommand\mc[1]{\multicolumn{1}{c}{#1}}
\usepackage{subfigure}
\addto\captionsenglish{}
\usepackage[colorlinks=true,linkcolor=blue,urlcolor=blue,citecolor=blue]{hyperref}
\begin{document}

\title{Virial coefficients of the Uniform Electron Gas from Path Integral Monte Carlo Simulations}
\author{G. R{\"o}pke}
\email{gerd.roepke@uni-rostock.de}
\affiliation{Institute of Physics, University of Rostock, 18059 Rostock, Germany}
\author{T. Dornheim}
\affiliation{Center for Advanced Systems Understanding (CASUS), 
Untermarkt 20, D-02826 G{\"o}rlitz, Germany}
\affiliation{Helmholtz-Zentrum Dresden-Rossendorf (HZDR), Bautzener Landstrasse 400, D-01328 Dresden, Germany}
\author{J. Vorberger}
\affiliation{Helmholtz-Zentrum Dresden-Rossendorf (HZDR), Bautzener Landstrasse 400, D-01328 Dresden, Germany}
\author{D. Blaschke}
\affiliation{Institute of Theoretical Physics, University of Wroclaw, 50-204 Wroclaw, Poland}
\affiliation{Center for Advanced Systems Understanding (CASUS), 
Untermarkt 20, D-02826 G{\"o}rlitz, Germany}
\affiliation{Helmholtz-Zentrum Dresden-Rossendorf (HZDR), Bautzener Landstrasse 400, D-01328 Dresden, Germany}
\author{B. Mahato}
\affiliation{Institute of Theoretical Physics, University of Wroclaw, 50-204 Wroclaw, Poland}

\date{\today}
\begin{abstract}
The properties of plasmas in the low-density limit are described by virial expansions. Analytical expressions are known from Green's function approaches only for the first three virial coefficients.
Accurate path integral Monte Carlo (PIMC) simulations have recently been performed for the uniform electron gas, allowing the virial expansions to be analyzed and interpolation formulas to be derived.
The exact expression for the second virial coefficient is used to test the accuracy of the PIMC simulations and the range of validity of the interpolation formula of Groth {\it et al.}~[Phys.~Rev.~Lett.~\textbf{119}, 135001 (2017)].
We discuss the fourth virial coefficient, which is of interest, e.g., for properties of solar plasmas, but has not yet been precisely known. 
Combining PIMC simulations with benchmarks from exact results of the virial expansion would allow us to obtain precise results for the equation of state (EoS) in a wide range of parameters. 

\end{abstract}

\maketitle

\section{Introduction}\label{sec0}

The thermodynamic properties of Coulomb systems in a wide region of density and temperature are of high interest with respect to various applications. A particularly important regime is given by so-called \emph{warm dense matter} (WDM)~\cite{wdm_book}, which naturally occurs in a gamut of astrophysical objects such as giant planet interiors~\cite{Benuzzi_Mounaix_2014} and brown dwarfs~\cite{becker}. Moreover, WDM plays an important role in technological applications such as the discovery and synthesis of materials~\cite{Kraus2016,Kraus2017,Lazicki2021}, hot-electron chemistry~\cite{Brongersma2015}, and inertial confinement fusion~\cite{hu_ICF}. As a result, WDM is actively realized in experiments at various research facilities such as the National Ignition Facility (NIF)~\cite{Moses_NIF}, the Linac Coherent Light Source (LCLS)~\cite{LCLS_2016}. and the Omega laser facility~\cite{Riley_2018} in the USA, or the European XFEL in Germany~\cite{Tschentscher_2017}; a topical overview of different relevant experimental techniques has been presented by Falk~\cite{falk_wdm}. At the same time, we stress that a rigorous theoretical description of such extreme states of matter is indispensable to interpret experimental measurements~\cite{kraus_xrts,Dornheim_T_2022}, and to guide the development of new set-ups~\cite{Dornheim_review,Hurricane_RMP_2023}.

In recent years, new possibilities to obtain results for the thermodynamic properties beyond perturbation theory have arisen, applying numerical simulations to solve the basic expressions~\cite{Dornheim_review,new_POP,review,wdm_book}.
Density-functional theory has been successfully applied to evaluate properties of warm dense matter~\cite{Bethkenhagen_PRR_2020,Ramakrishna_PRB_2021,Moldabekov2022,Karasiev_PRE_2022}, but a main deficit of it is that electron-electron interaction is treated in a certain approximation~\cite{JCTC2023,moldabekov2023bound}.
Therefore, computationally more involved path integral Monte Carlo (PIMC) simulations~\cite{cep,Militzer_PRE_2001,review,Dornheim_HEDP_2022} are of growing interest since they allow the correct treatment of electron-electron interaction.

As a simple example, we consider the homogeneous electron gas (uniform electron gas, UEG~\cite{review,quantum_theory,loos}), where the electrons move over a positively charged background which is added to ensure charge neutrality. 
The electronic part of the Hamiltonian is  given by
\begin{equation}
\label{Hamilt}
\hat{H}=\sum_i^N\frac{ \hat{\bf p}_i^2}{2 m}+\frac{1}{2}\sum_{i \neq j}\frac{e^2}{4 \pi \epsilon_0 |{\bf \hat{r}}_i-{\bf \hat{r}}_j|}\quad ,
\end{equation}
 where $m$ and $e$ are the electron mass and charge,  $\epsilon_0$ is the permittivity of the vacuum, and $\hat{\bf p}_i$ and $\hat{\bf r}_i$ denote the momentum and position operators of the $i$-th electron.

In thermodynamic equilibrium, the state of the plasma is determined by the temperature $T$ in addition to the number density $n=N/\Omega$ (with $\Omega$ being the volume) or the corresponding chemical potential $\mu$.  Note that we consider the unpolarised UEG throughout, where $n_\uparrow = n_\downarrow=n/2$ so that both spin directions have the same density.
The relationships between the various state variables such as internal energy $U$, free energy $F$, entropy $S$, pressure $P$, etc., are called equations of state (EoS). All thermodynamic properties can be derived from a thermodynamic potential; $F(\Omega, N,T)$ as function of $\Omega, N,T$  constitutes an example. We note that EoS databases constitute key input for a host of practical applications such as the modeling of laser fusion~\cite{hu_ICF} or the description of astrophysical objects~\cite{saumon1,becker,wdm_book}. 

Correlations appear for the plasma owing to the Coulomb interaction term (\ref{Hamilt}) proportional to $e^2$. No closed-form solutions are known, and we must perform approximations (or use numerical techniques~\cite{Dornheim_review}) to solve this many-body problem.
We discuss two possibilities:

(i) Perturbation expansion with respect to $e^2$. We obtain analytic expressions for arbitrary orders of $e^2$ in terms of noninteracting equilibrium correlation functions, 
which can be easily evaluated using Wick's theorem. However, we have no proof of the convergence of this series expansion and no error estimate.
In order to make this analytical approach more efficient, the methods of thermodynamic Green's functions and Feynman diagram technique 
were elaborated \cite{FW,KKER}.
The perturbation approach is improved by performing partial summations corresponding to special concepts such as the introduction of the quasiparticle picture (self-energy $\Sigma$),
 screening of the potential (polarization function $\Pi$), or formation of bound states (Bethe-Salpeter equation). 
This leads to useful results for the properties of the plasma in a wide range of $T$ and $n$. 
However, as characteristic for perturbative approaches, exact results can be found only in some limiting cases.
 
 (ii) In principle, an accurate evaluation of thermodynamic potentials is possible using path-integral Monte Carlo (PIMC) simulations, see~Refs.~\cite{review,Dornheim_PRE_2020,new_POP} and references therein.
The shortcomings of this approach include the relatively small number of particles (a few dozen up to one hundred at the present time~\cite{dornheim_prl,Dornheim_JCP_2021}) and the sign problem for fermions~\cite{dornheim_sign_problem,Dornheim_2021}.
Over recent years, this emerging approach has been put forward together with improving computer facilities.
At present, accurate calculations have been performed mainly for the UEG over a broad range of parameters~\cite{Dornheim_HEDP_2022,dornheim_electron_liquid}.

The UEG is the simplest example. In a next step, calculations for the two-component hydrogen plasma would be of interest for both thermodynamics and transport properties~\cite{Militzer_PRE_2001,Bohme_PRL_2022,filinov2023equation,Hamann_PRR_2023}.
 There are some low-density results, see Militzer~\cite{Militzer_PRE_2001} or Filinov and Bonitz~\cite{filinov2023equation} and further references given in these works.
However, high-precision PIMC simulations for hydrogen plasmas in the low-density region, which allow the extraction of higher order virial coefficients, are presently  not available. 

In this work, we investigate in detail exact virial expansions, which are of considerable value as a rigorous benchmark for numerical methods in certain limits, and as a useful constraint for (semi-)analytical EoS interpolations.
To this end, we present new PIMC simulations for the UEG under extreme conditions, i.e. at very low densities and very high temperatures.
We investigate the virial expansion and discuss higher order virial coefficients not considered in previous publications~\cite{Dornheim_HEDP_2022}. In particular, we discuss the high temperature limit of the fourth virial coefficient. A comparison is made with the interpolation formula \cite{groth_prl} and the limits of its applicability are shown.

The paper is organized as follows: 
A brief introduction to the virial expansion of the mean potential energy is given in Sec.~\ref{Sec:1}. 
Effective virial coefficients and virial plots are introduced. 
PIMC simulations at high temperatures and small densities are presented in Sec.~\ref{Sec:2}. 
An interpolation formula~\cite{groth_prl}  is shown in Sec.~\ref{Sec:3}, and the second virial coefficient is considered as a benchmark. 
The fourth virial coefficient is analysed in Sec.~\ref{Sec:4}, where a high-temperature approximation is given and compared with PIMC simulations. 
The exact temperature dependence of the fourth virial coefficient is not yet known, but remains a challenge for future PIMC simulations, as 
we conclude in Sec.~\ref{sec7}.

\section{Virial coefficients from analytical approaches}\label{Sec:1}

\subsection{Virial expansions for the UEG}

Using the method of thermodynamic Green's functions from quantum statistics,
 the virial expansion of the free energy of the UEG is written as
\begin{multline}
\label{Fvir}
F(T,\Omega,N)=\Omega k_BT \Big\{n \ln n + \left[\frac{3}{2}\ln\left(\frac{2 \pi \hbar^2}{m k_BT}\right)-1\right] n\\
-F_0(T)n^{3/2}-F_1(T)n^2 \ln n-F_2(T) n^2\\
-F_3(T)n^{5/2} \ln n
-F_4(T) n^{5/2}+{\cal O}(n^3\ln n)\Big\},
\end{multline}
see  Refs.~\cite{KKER,Dornheim_HEDP_2022} where expressions for the lowest virial coefficients $F_i$ are also given.

The mean potential energy $V$ is  given by
\begin{equation}
V(T,\Omega,N)=e^2 \frac{\partial}{\partial (e^2)} F(T,\Omega,N)\,,
\end{equation}
(for the relation to the internal energy see Ref.~\cite{Kraeft02}). 

From the virial expansion of  $F(T,\Omega,N) $, see Ref.~\cite{KKER}, we get the following virial expansion of $V$
\begin{widetext}
\begin{multline}
\frac{V}{Nk_{\rm B}T}=-\frac{\kappa^3}{8 \pi n}- \pi n\lambda^3 \tau^3 \ln(\kappa\lambda)
-\pi n \lambda^3\left[\frac{\tau}{2}-\frac{\sqrt{\pi}}{2} (1+\ln(2))\tau^2+\left(\frac{C}{2}+\ln(3)-\frac{1}{3}+\frac{\pi^2}{24}\right)\tau^3\right. \\
\left. +\sqrt{\pi}\sum_{m=4}^\infty\frac{(-1)^mm}{2^m \Gamma(m/2+1)}\left[2 \zeta(m-2)-(1-4/2^m)\zeta(m-1)\right]\tau^m \right] 
- \pi n\lambda^4 \tau^4 \kappa \ln(\kappa\lambda)+\frac{V_4(T)}{Nk_{\rm B}T}n^{3/2} +{\cal O}(n^2 \ln(n))
\label{virV}
\end{multline}
\end{widetext}
with the variables
\begin{equation}
\kappa^2 = \frac{ne^2}{\epsilon_0k_{\rm B}T},\qquad \lambda^2=\frac{\hbar^2}{m k_{\rm B}T}, \qquad \tau = \frac{e^2 \sqrt{m}}{4 \pi \epsilon_0 \sqrt{k_{\rm B}T}\hbar}.
\end{equation}
Here $\zeta(x)$ denotes the Riemann zeta function, and $C=0.57721\dots$ is Euler's constant. We express this expansion in terms of $T, n$ and introduce atomic units $\hbar=m=e^2/4 \pi \epsilon _0=1$ (see Appendix \ref{App:1}),
so that $k_{\rm B}T$ is measured in Hartree (Ha) and $n$ in electrons per $a_{\rm B}^3$, $n_{\rm B}=n a_{\rm B}^3$.

The virial expansion of the specific mean potential energy $v=V/N$ is as follows ($\kappa^2 \lambda^2=4 \pi n_{\rm B}/T^2_{\rm Ha}$)
\begin{multline}
\label{virialexp}
v(T,n)=v_0(T) n_{\rm B}^{1/2}+v_1(T) n_{\rm B} \ln\left(\kappa^2 \lambda^2\right)+v_2(T) n_{\rm B}
\\
+v_3(T)n_{\rm B}^{3/2} \ln\left(\kappa^2 \lambda^2\right)+v_4(T) n_{\rm B}^{3/2}+{\cal O}(n^2 \ln(n))\,.
\end{multline}
with
\begin{widetext}
\begin{align}
v_0(T)=&-\frac{\sqrt{\pi}}{T^{1/2}_{\rm Ha}},\nonumber\\ 
v_1(T)=&-\frac{\pi}{2 T_{\rm Ha}^2},\nonumber\\
v_2(T)=&-\frac{\pi}{T_{\rm Ha}}\left[\frac{1}{2}-\frac{\sqrt{\pi}}{2}(1+\ln(2))\frac{1}{T_{\rm Ha}^{1/2}}+\left(\frac{C}{2}+\ln(3)-\frac{1}{3}+\frac{\pi^2}{24} \right) \frac{1}{T_{\rm Ha}}\right.\nonumber \\
&\left. - \sqrt{\pi}\sum_{m=4}^\infty\frac{m}{2^m \Gamma(m/2+1)}\left(\frac{-1}{T_{\rm Ha}^{1/2}}\right)^{m-1} [2 \zeta(m-2)-(1-4/2^m)\zeta(m-1)]\right],\nonumber \\
v_3(T)=&-\frac{3 \pi^{3/2}}{2 T_{\rm Ha}^{7/2}}.
\label{v0123}
\end{align}
\end{widetext}
No closed expression for $v_4(T)$ is known.

\subsection{The effective second virial coefficient}

In Refs.~\cite{Dornheim_HEDP_2022,R23}, a method for extracting the virial coefficients from data was presented.
We demonstrate this approach for the second virial coefficient $v_2(T)$ for which the exact expression (\ref{v0123}) is known.
Since in the low-density limit the lowest virial coefficients dominate the function $v(T,n)$ (\ref{virialexp}), we subtract the "trivial" contributions of $v_0(T)$ (Debye term) and $v_1(T)$.
The remaining part is then dominated by $v_2(T)$ in the low-density limit.

To extract the value of $v_2(T)$ from numerical (or measured) results for $v(T,n)$, we consider isotherms and calculate an effective, density-dependent second virial coefficient
\begin{multline}
\label{v2eff}
v_2^{\rm eff}(T,n)=\left[v(T,n)-v_0(T)n_{\rm B}^{1/2}\right.\\
\left.-v_1(T)n_{\rm B} \ln\left(\frac{4 \pi n_{\rm B}}{T_{\rm Ha}^2}\right)\right]/n_{\rm B}.
\end{multline}
We have the result $v_2(T) = \lim_{n \to 0} v_2^{\rm eff}(T,n)$. 
The density dependence of $v_2^{\rm eff}(T,n)$ in the low-density limit is given according to Eq.~(\ref{virialexp}) as
\begin{equation}
\label{v2eff1}
v_2^{\rm eff}(T,n)=v_2(T)+v_3(T)n_{\rm B}^{1/2} \ln(4 \pi n_{\rm B}/T_{\rm Ha}^2)+{\cal O}[n^{1/2}].
\end{equation}
So in the virial plot where $v_2^{\rm eff}(T,n)$ is plotted as a function of $n_{\rm B}^{1/2} \ln(4 \pi n_{\rm B}/T_{\rm Ha}^2)$, the isotherms should meet the co-ordinate at $v_2(T)$ and the slope is $v_3(T)$.
The linear pattern is violated when higher virial coefficients become relevant. Note that both $v_2(T)$ and $v_3(T)$ are known for the UEG according to Eq. (\ref{v0123}).
Corresponding plots for three isotherms are shown below in Figs. \ref{fig:isotherm1}-\ref{fig:3}, $T_{\rm Ha}=0.589307, \,\,0.294653$, and 100.

We will apply this method of the virial plot to PIMC simulations $v^{\rm PIMC}(T,n)$ to obtain the values $v_2^{\rm eff,PIMC}(T,n)$ according to Eq. (\ref{v2eff}).
It is clear that this method of extracting virial coefficients requires a high precision of the calculated data, since we are analysing the difference of large numbers. 
This is because the lower virial coefficients, such as the Debye term, dominate the low-density limit of the potential energy density $v(T,n)$.

\section{PIMC simulations for the UEG}
\label{Sec:2}

\subsection{PIMC simulations at high temperatures and small densities}


\begin{figure*}
\includegraphics[width=0.462\textwidth]{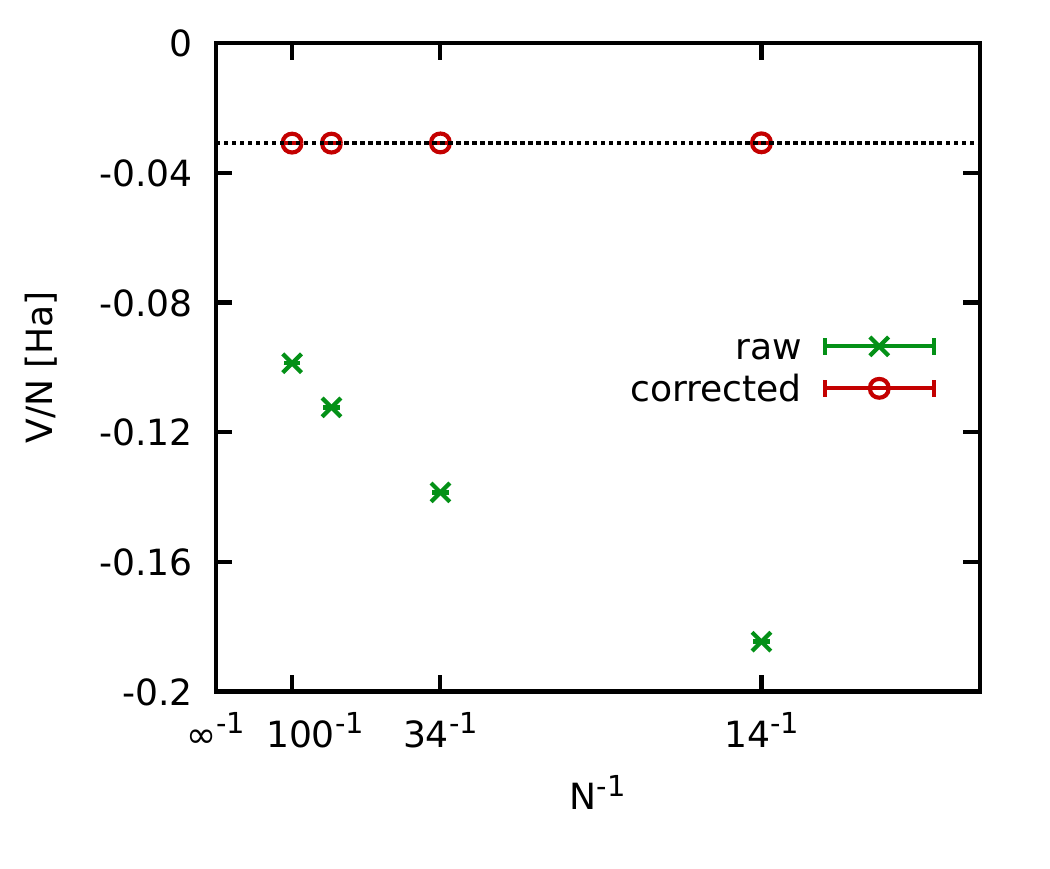}\includegraphics[width=0.462\textwidth]{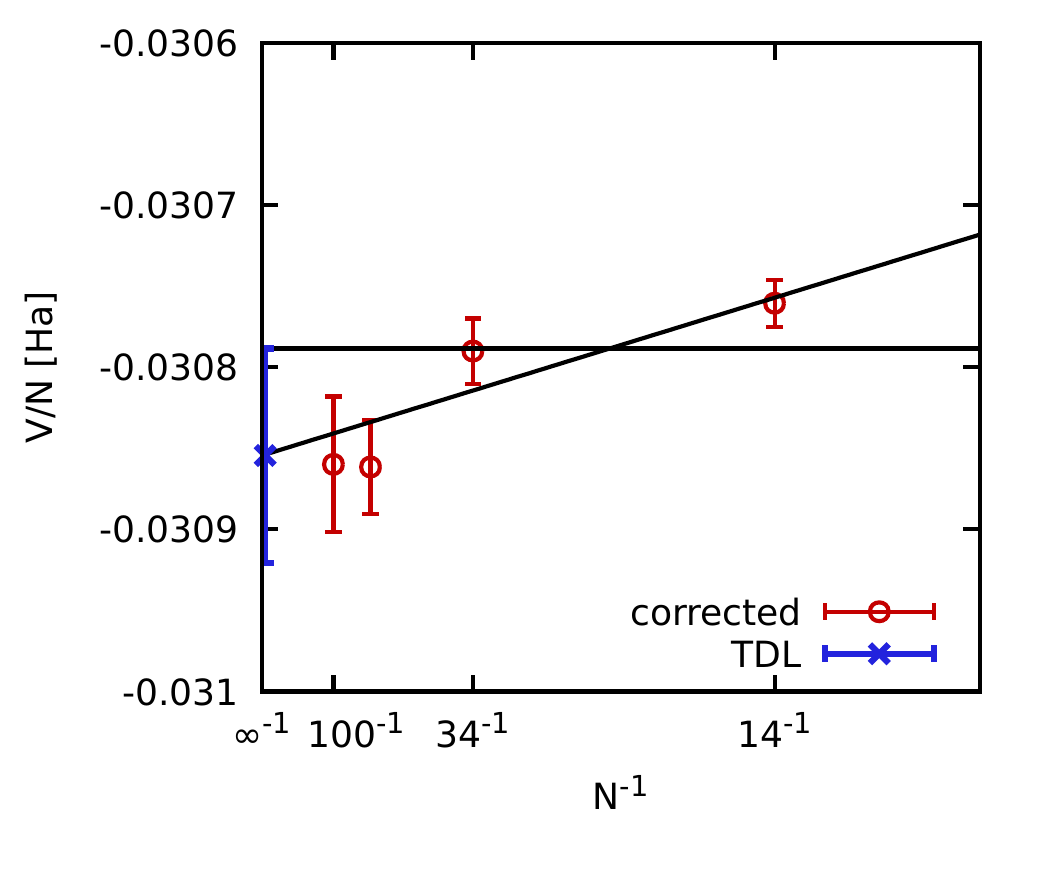}
\caption{\label{fig:Extrapolated_UEG_data}
Extrapolating PIMC results for the interaction energy of the UEG to the thermodynamic limit (TDL) at $r_s=2$ and $\Theta=217.204$ ($T_\textnormal{Ha}=100$). Left: Raw PIMC results for the interaction energy per particle $V/N$ (green crosses), and finite-size corrected values (red circles) as a function of system size. Right: Magnified segment around the finite-size corrected results; the solid black lines show empirical linear and constant fits, and the blue cross depicts the extrapolated result in the limit of $N\to\infty$.
}
\end{figure*} 

To compare with the virial expansion, we have to consider high temperatures $T_{\rm Ha} \ge 1$ and $n_{\rm B}^{1/2} \ln(4 \pi n_{\rm B}/T_{\rm Ha}^2) \ll 1$ so that the contributions of higher order virial coefficients are sufficiently small.
This requires us to go beyond the conditions that had been explored
in Ref.~\cite{Dornheim_HEDP_2022}. Clearly, the fermion sign problem~\cite{dornheim_sign_problem} does not pose an obstacle as quantum degeneracy effects become negligible in the limit of high temperature. Instead, the biggest challenge is given by finite-size effects in the simulation data, which are substantial in this regime. This is illustrated in the left panel of Fig.~\ref{fig:Extrapolated_UEG_data}, where we show raw PIMC simulation results for the interaction energy per particle $V/N$ at $r_s=2$ and $\Theta=217.204$ (for $r_s, \Theta$, see Appendix \ref{App:1}).
Indeed, the dependence on the system size is of the order of $100\%$.
To overcome this bottleneck, we employ the finite-size correction scheme developed in Ref.~\cite{dornheim_prl}, which constitutes a finite-temperature version of the approach originally introduced by Chiesa \textit{et al.}~\cite{Chiesa_PRL_2006}. We refer the interested reader to the overview~\cite{review} for a more detailed discussion.
The thus corrected interaction energies are shown as the red circles in Fig.~\ref{fig:Extrapolated_UEG_data} and exhibit a drastically reduced dependence on the system size; the residual error is of the order of $\sim0.1\%$. In the right panel, we show a magnified segment around these corrected values, and the solid black lines show empirical fits based on simple linear and constant functional forms. In practice, we give the final result based on the linear extrapolation, and the associated uncertainty is computed from the difference between the two solid lines, see the blue cross in Fig.~\ref{fig:Extrapolated_UEG_data}. All PIMC results shown in this work have been obtained based on this procedure.

\subsection{Calculations with PIMC simulation results}

We show  PIMC simulation results for the UEG as published in Ref.~\cite{Dornheim_HEDP_2022} in Tab.~\ref{tab:1}.
For our analysis, we need high accuracy data because we consider small differences of large numbers. 
We focus only on results for $r_s = 20$ and $\Theta =128$ and $\Theta =64$. 
There, the virial coefficients are relatively large, and higher orders of the virial expansion are not too dominant.

\begin{figure*}[t]
\centering
\subfigure[\label{fig:v3eff100}]{\includegraphics[width=0.49 \textwidth]{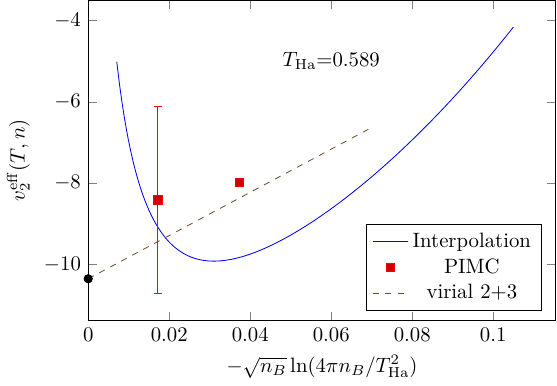}}\hfill
\subfigure[\label{fig:v2vsn05}]{\includegraphics[width=0.49 \textwidth]{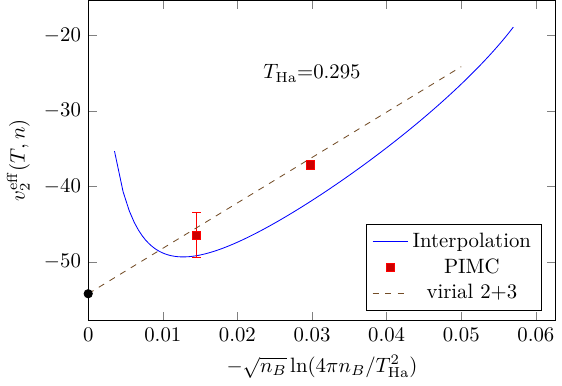}}
\caption{Isotherms for $T_{\rm Ha} = 0.589$ and  $0.295$. In both subfigures the effective second virial coefficients $v_2^{\rm eff, GDSMFB}(T,n)$, Eq. (\ref{v2eff}), are plotted as function of $-n_{\rm B}^{1/2}\ln(4 \pi n_{\rm B}/T_{\rm Ha}^2)$. The exact contribution of $v_2(T)$ and $v_3(T)$ is given by the dashed line [virial 2+3: Eq. (\ref{v2eff1})]. The red squares corresponds to $v_2^{\rm eff, PIMC}(T,n)$. See Tabs.  \ref{tab:1} and \ref{tab:2} (Atomic units used.)
}
\label{fig:isotherm1}
\end{figure*}

The virial plots for the corresponding two isotherms are presented in Fig.~\ref{fig:isotherm1}, where $v_2^{\rm eff, PIMC}(T,n)$ is shown as function of $n_{\rm B}^{1/2} \ln(4 \pi n_{\rm B}/T_{\rm Ha}^2) $.  Isotherms  are presented because the virial coefficients describe the expansion with respect to density $n$ at fixed $T$.
For comparison, the benchmarks $v_2(T)+v_3(T) n_{\rm B}^{1/2} \ln(4 \pi n_{\rm B}/T_{\rm Ha}^2) $  are also shown. There is a nice agreement. Deviations may be explained by the contribution of higher virial coefficients for the analytical results. In addition, the PIMC data have also uncertainties expressed by error bars.

\begin{table*}[htp]
\caption{PIMC data from Ref. \cite{Dornheim_HEDP_2022,R23}, corresponding parameter values and effective second virial coefficients (atomic units, see Appendix \ref{App:1}), }
\begin{center}
 \begin{tabular}{lccccccS[table-format=2.4]}
\toprule
No.&$r_s$& $\Theta$  &  $T_{\rm Ha}$& $n_{\rm B}$ & $v^{\rm PIMC}$[Ha] 
& $n_{\rm B}^{1/2} \ln(4 \pi n_{\rm B}/T_{\rm Ha}^2) $ & {$v_2^{\rm eff, PIMC}$[Ha]} \\
\hline
1 &20    & 128    &0.589307    & 0.0000298416    & -0.0119299   
& 0.0373158  & -7.9796    \\
2& 20    & 64        &0.294653    & 0.0000298416    & -0.0160051    
    & 0.0297429  & -37.1054      \\
    \toprule
 \end{tabular}
\end{center}
\label{tab:1}
\end{table*}

To demonstrate the limiting behavior given in the virial plot by the linear relation (\ref{v2eff1}), neglecting higher order terms ${\cal O}[n^{1/2}]$, more PIMC simulation data would be of interest. 
In this work, we performed high precision PIMC simulations  for additional three  parameter values. The results are shown in Tab. \ref{tab:2}.
\begin{table*}[htp]
\sisetup{separate-uncertainty}
\caption{New PIMC data, corresponding parameter values and effective second virial coefficients (atomic units, see Appendix \ref{App:1}). }
\begin{center}
 \begin{tabular}{l S[table-format=2]S[table-format=3.3]S[table-format=3.6]S[table-format=1.4e-1]S[table-format=-1.8+-1.8]S[table-format=1.5]S[table-format=-2.7]}
\toprule
\mc{No.} &\mc{$r_s$}& \mc{$\Theta$}  &  \mc{$T_{\rm Ha}$}& \mc{$n_{\rm B}$} & \mc{$v^{\rm PIMC}$[Ha]} & \mc{$n_{\rm B}^{1/2} \ln(4 \pi n_{\rm B}/T_{\rm Ha}^2) $} & \mc{$v_2^{\rm eff, PIMC}$[Ha]} \\
\hline
1 & 40    & 512    &0.589307    & 3.7301e-6   & -0.00434040 +- 0.0000086  & 0.01721& -8.4189922     \\
2 & 40 & 256        &0.294653    & 3.7301e-6   & -0.00597188 +- 0.00001116  & 0.01453  & -46.4374891    \\
3 & 2 & 217.2        &100    & 2.9842e-2     & -0.03085446 +- 0.00006612  & 1.76049  & -0.0095033     \\
    \toprule
 \end{tabular}
\end{center}
\label{tab:2}
\end{table*}

In Tab. \ref{tab:2}, No. 1 and 2 belong to the isotherms of Tab. \ref{tab:1}, No. 1 and 2, but at lower density. 
As seen in Fig.  \ref{fig:isotherm1}, the PIMC simulations are consistent with the virial expansion. 
However, the error bars are quite large so that the extrapolation $n \to 0$ to extract the second virial coefficient from PIMC simulations has also a large error.

In Tab. \ref{tab:2}, No. 3 belongs to the isotherm $T_{\rm Ha}=100$ to study the high-temperature limit. As shown in Fig.  \ref{fig:3}, the PIMC simulation data are also consistent with the virial benchmark. However, the error bars are large, too.

\section{Interpolation formulas for thermodynamic properties of the UEG}
\label{Sec:3}

The PIMC simulations are computationally very expensive. 
Instead of performing time-consuming calculations for each parameter value, interpolation formulas have been worked out  which allow to reproduce the results for each parameter value within a given accuracy. 
Because the limiting behavior of the free energy is known at low and at high density, Pad\'e expressions can be used to good effect~\cite{GDB2017,ksdt,groth_prl}.
We can also test these interpolation formulas with respect to their accuracy and the parameter range where they can be used.
In particular, we study whether they can be used instead of PIMC simulations to extract a value for the virial coefficient such as $v_2(T)$ or $v_4(T)$.

The GDSMFB interpolation formula for the XC free energy density of the spin-unpolarized UEG is~\cite{groth_prl} 
\begin{equation}
f^{\rm GDSMFB}_{\rm XC}(r_S,\Theta)=-\frac{1}{r_s}\, \frac{a(\Theta)+ b(\Theta) \sqrt{r_s}+c(\Theta)r_s}{1+d(\Theta) \sqrt{r_s}+e(\Theta)r_s}.
\end{equation}
The coefficients $a,b,c,d,e$ are again Pad\'e formulae with respect to temperature given in the Supplemental material to  \cite{groth_prl} [see also Appendix \ref{sec:Parameter}].

The exact relationship between the exchange-correlation free energy and the potential energy is given as
\begin{equation}
f_{\rm XC}(r_s,\Theta)=\frac{1}{r_s^2} \int_0^{r_s}dr'_s\,r'_s v(r'_s,\Theta)\,,
\end{equation}
so that 
\begin{equation}
v(r_s,\Theta)=2 f_{\rm XC}(r_s,\Theta)+r_s \frac{\partial f_{\rm XC}(r_s,\Theta)}{\partial r_s} {\Big{|}}_\Theta\,.
\end{equation}
We find

\begin{figure}
\centerline{\includegraphics[width=0.49 \textwidth]{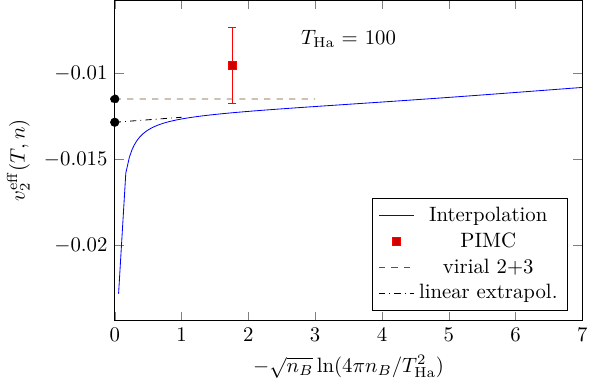}}
\caption{Effective second virial coefficient $v_2^{\rm eff, GDSMFB}(T,n)$, Eq. (\ref{v2eff}), 
plotted as function of $-n_{\rm B}^{1/2}\ln(4 \pi n_{\rm B}/T_{\rm Ha}^2)$. 
The exact value $v_2(100)=-0.0114705$ is also shown  [virial 2+3: Eq. (\ref{v2eff1})]. The slope according to $v_3$ becomes very small at high temperatures. In addition a PIMC simulation according Tab. \ref{tab:2}, No. 3, is presented. A linear extrapolation $ - 0.0128 + 0.000289 \sqrt{n_{\rm B}}\log(4\pi n_{\rm B}/T_{\rm Ha}^2)$ is also shown (dashdotted line).
(Atomic units used.) \label{fig:3}}
\end{figure}

\begin{widetext}
\begin{equation}
\label{GDSMFB}
v^{\rm GDSMFB}_{\rm XC}(r_S,\Theta)=-\frac{1}{r_s}\, \frac{a+ b \sqrt{r_s}+c \,r_s}{1+d \sqrt{r_s}+e\,r_s}+\frac{1}{r_s}\, \frac{[(ad-b)/2]\sqrt{r_s}+(ae-c) r_s+[(be-cd)/2]\, r_s^{3/2}}{(1+d \sqrt{r_s}+e\,r_s)^2}.
\end{equation}
\end{widetext}
This expression will be used to calculate $v_2^{\rm eff, GDSMFB}(T,n)$ according to Eq.~(\ref{v2eff}).
The corresponding results are shown in  Figs.~\ref{fig:isotherm1} and \ref{fig:3} as blue lines.

It is obvious that the interpolation formula shows strong deviations in the low-density limit. 
The reason is that the Pad\'e formula (\ref{GDSMFB}) is not constructed to reproduce the $v_2(T)$ so that the analytical behavior of the Pad\'e formula in the low-density limit is not consistent with the virial expansion.
This discrepancy shows a limit of applicability of the interpolation formula. 
However, because in the low-density region the lowest order virial terms (e.g. the Debye shift) dominate, the error of the interpolation formula becomes small if these lowest orders are correctly included.

Is it possible to extract the virial coefficients from the Pad\'e formula (\ref{GDSMFB})? 
In Fig. \ref{fig:3} we show the values for the effective second virial coefficient $v_2^{\rm eff, GDSMFB}(T,n)$, Eq. (\ref{v2eff}), for the isotherm $T_{\rm Ha}=100$. 
It is clearly shown that in the low-density limit (below $n_{\rm B}^{1/2}\ln(4 \pi n_{\rm B}/T_{\rm Ha}^2)=1$) the benchmark of the second virial coefficient is not reached.
It  is also shown that there the value of the interpolation formula lies outside the error bars of the PIMC simulation.

We see that in the limit $n \to 0$ the results for $v_2^{\rm eff, GDSMFB}(T_{\rm Ha}=100,n)$ will not match the exact value $v_2(100)=-0.0114705$. 
Because the Pad\'e interpolation formula did not reproduce the second virial coefficient, strong deviations become dominant near $n=0$. 
A linear extrapolation of the values for $n_{\rm B}^{1/2}\ln(4 \pi n_{\rm B}/T_{\rm Ha}^2)>2$ gives a limit at $n=0$ of -0.0128 which deviates from the exact value by about 10 \%. 

In conclusion, from Figs. \ref{fig:isotherm1} \& \ref{fig:3} we see that PIMC simulations become difficult in the low-density region and the error bars become large. The low-order virial coefficients may be considered as a benchmark for the simulation.
The interpolation formula (\ref{GDSMFB}) describes the virial plot for $v_2^{\rm eff}$ in a certain approximation only in an intermediate parameter range.
At high densities, higher orders of the virial expansion become important.
At very low densities, the analytical behavior of the interpolation formula is not able to reproduce the exact virial coefficients. 
However, they may be estimated in certain approximation, if a linear behavior can be seen in the virial plot, see Fig. \ref{fig:3}.

\section{The fourth virial coefficient}
\label{Sec:4}

\begin{figure*}[t]
\centering
\subfigure[\label{fig:v3eff100}]{\includegraphics[width=0.49 \textwidth]{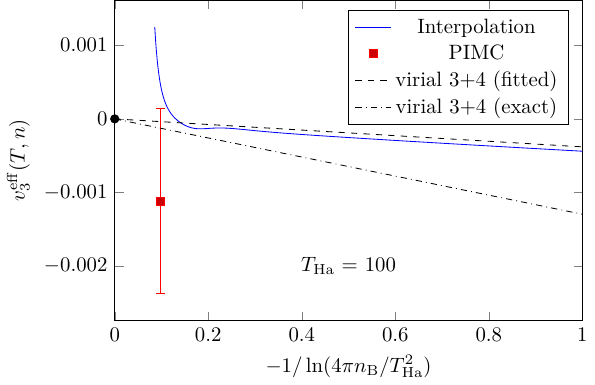}}\hfill
\subfigure[\label{fig:v2vsn05}]{\includegraphics[width=0.49 \textwidth]{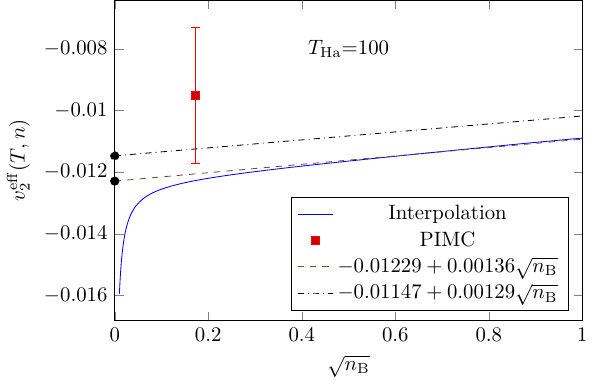}}
\caption{ Isotherms for $T_{\rm Ha}=100$. (a) The effective third virial coefficient $v_3^{\rm eff, GDSMFB}(T,n)$, Eq. (\ref{v3eff}), plotted as a function of $-1/\ln(4 \pi n_{\rm B}/T_{\rm Ha}^2)$. The slope of $v_3^{\rm eff}(T,n)$ determines $v_4(100)$. The linear relation Eq. (\ref{v3eff}) with $v_3(100)=-8.352 \times 10^{-7}$ is denoted as virial 3+4. The dashed curve corresponds to $v^{\rm GDSMFB}_4=0.00038$. The dash-dotted curve corresponds to $v_4=0.00129$, Eq. (\ref{v4approx}).\,\, 
(b) The effective second virial coefficient $v_2^{\rm eff, GDSMFB}(T,n)$, Eq. (\ref{v2eff}), 
plotted as function of $n_{\rm B}^{1/2}$. From analytical approaches (Eqs. \eqref{v0123} and \eqref{v4approx}) follows $v^{\rm eff}_2(100)\approx -0.01147+0.00129 n_{\rm B}^{1/2}$ (shown as dash-dotted line). The dashed line represents a linear fit. 
(Atomic units used.) 
}
\label{fig:isotherm100}
\end{figure*}

\subsection{High-temperature limit and PIMC simulation data}
Analytical expressions for $v_4(T)$ are not yet known. Approximations considering special classes of diagrams have been obtained within Green's function approaches. For instance, considering the diagrams of lowest order with respect to interaction, in Refs. \cite{KKER,Riemann95}, the following contribution to the fourth virial coefficient has been given [in atomic units, see Eqs. (\ref{virialexp}), (\ref{v0123})]
\begin{equation} 
\label{v4approx}
    v_4(T) \approx \frac{3}{2}\frac{\pi^2}{T_{\rm Ha}^2}-\frac{10}{3}\frac{\pi^{3/2}}{T_{\rm Ha}^{5/2}}.
\end{equation}
This result leads to a high temperature behavior $\lim_{T \to \infty} v_4(T) \propto T^{-2}$ if no other diagrams contribute to this limit.

We follow the method explained for the second virial coefficient.
At fixed $T$, in the low-density limit the lowest virial coefficient will dominate because of the analytical behavior near $n=0$. 
Thus, subtracting the $(k-1)$ lowest virial coefficients from the thermodynamic quantity,
the remaining part allows to determine the next virial coefficient $v_k(T)$ \cite{R23}. 
We use a virial plot for $v_3^{\rm eff}(T,n)$ defined as 
\begin{multline}
\label{v3eff}
v_3^{\rm eff}(T,n)=\left[v(T,n)-v_0(T)n_{\rm B}^{1/2}-v_1(T)n_{\rm B} \ln\left(\frac{4 \pi n_{\rm B}}{T_{\rm Ha}^2}\right)\right.\\
-v_2(T) n_{\rm B} \Big]/[n_{\rm B}^{3/2} \ln(4 \pi n_{\rm B}/T_{\rm Ha}^2)] .
\end{multline}
In the low-density limit, the density dependence of $v_3^{\rm eff}(T,n)$  is given according to Eq. (\ref{virialexp}) as
\begin{equation}
\label{v3eff1}
v_3^{\rm eff}(T,n)=v_3(T)+v_4(T)\frac{1}{\ln(4 \pi n_{\rm B}/T_{\rm Ha}^2)}+{\cal O}[n^{1/2}].
\end{equation}
Thus, in the virial plot where $v_3^{\rm eff}(T,n)$ is shown as a function of $1/ \ln(4 \pi n_{\rm B}/T_{\rm Ha}^2)$, isotherms should meet the co-ordinate at $v_3(T)$, and the slope is $v_4(T)$.

We discuss here the high-temperature region because we expect that the odd virial coefficients become small in this limit $T \to \infty$, as seen for the lowest virial coefficients (\ref{v0123}), see also Tab. \ref{tab:4} in App. \ref{app:3}. 
We expect a wider range of the linear relation (\ref{v3eff1}) if the higher virial coefficients, in particular $v_5(T)$, are small. 
As example, we consider $T_{\rm Ha}=100$, see Fig. \ref{fig:v3eff100}. 
The dash-dotted curve denotes the virial expansion with the exact value for $v_3(100)$, Eq. (\ref{v0123}), and the approximation (\ref{v4approx}) for $v_4(100)$. In addition, a PIMC simulation is also shown (No. 3 from Tab. \ref{tab:2}). Within the error bars, the result agrees with the virial expansion.
However, for this approach to extract virial coefficients from PIMC simulations, more data in the low-density region with higher accuracy are required which are not yet available.

\subsection{Fourth virial coefficient from interpolation formulas}

Because the full $T$ dependence of $v_4(T)$ is not yet known from the Green's function approach,
it would be of interest to obtain results from simulations. 
As shown for $v_2(T)$ in the previous section, see also \cite{Dornheim_HEDP_2022}, high-accurate PIMC simulations may be used to extract this quantity. However, they are not yet available.

To make some estimations with respect to $v_3(T)$, we may use  the interpolation formula (\ref{GDSMFB}) instead of the exact approach using PIMC simulations. 
Because this GDSMFB interpolation formula is only an approximation, significant deviations may occur.

Using $v^{\rm GDSMFB}(T,n)$  (\ref{GDSMFB}) as input for $v$, we calculate $v_3^{\rm eff,GDSMFB}(T,n)$ according Eq. (\ref{v3eff}), see Appendix \ref{app:3}.
These values are plotted in Fig. \ref{fig:v3eff100} as a function of $-1/\ln(4 \pi n_{\rm B}/T_{\rm Ha}^2)$.
As discussed above, in the high temperature region considered here, the odd virial coefficients give only small contributions.

From this curve, the linear extrapolation is possible for the values at larger densities. The value of $v_4(100)$ is estimated to be close to 0.00038.
The values at smaller densities cannot be used for the extrapolation because the interpolation formula is an approximation, 
and deviations yield large effects for small densities as already seen for $v_2^{\rm eff,GDSMFB}(T,n)$ in Figs. \ref{fig:isotherm1} and \ref{fig:3}.

In Fig. \ref{fig:v3eff100}, we also show the approximation (\ref{v4approx}) with the value $v_4(T_{\rm Ha}=100) \approx 0.001295$. Compared with the PIMC value of Tab. \ref{tab:2} also shown in Fig. \ref{fig:v3eff100}, we find that both results are consistent. The GDSMFB interpolation formula is not consistent with PIMC simulations in this parameter region. In addition, the extracted approximation for $v_4(100)$ is different from the approximation (\ref{v4approx}) which should be valid in the high temperature limit.

\subsection{Generalized virial plots}

The virial plots use an abscissa which gives a linear relation for the next higher virial coefficient so that this next virial coefficient is extracted from the slope of the isotherms at zero density.
It may happen that the virial expansion contains terms which are very small so that these terms are not relevant. 
For instance, at high temperatures, $v_3(T)$ becomes very small because it behaves $\propto T^{-7/2}$ according Eq. (\ref{v0123}). 
As shown in Fig. \ref{fig:3}, the slope of virial 2+3 is near to zero, but the interpolation formula indicates a significant increase with density.
This is the contribution of higher order virial coefficients.
We assume that in the high-temperature limit, the virial term $v_3(T)$ can be neglected so that the dominant contribution to the virial expansion in the low-density range follows from $v_4(T)$.

To extract this leading virial coefficient $v_4(T)$ from $v_2^{\rm eff}(T,n)$, Eq. \eqref{v2eff1},
\begin{multline}
\label{v2eff2}
v_2^{\rm eff}(T,n)=v_2(T)+v_3(T)n_{\rm B}^{1/2} \ln(4 \pi n_{\rm B}/T_{\rm Ha}^2)+v_4(T)n_{\rm B}^{1/2}\\
+{\cal O}[n^{1/2}\ln(4 \pi n/T^2)]
\end{multline}
we introduce a generalized virial plot where the abscissa is $n^{1/2}$. If we observe a linear behavior, the slope determines $v_4(T)$. 
The third virial coefficient gives a contribution only for very low densities, leading to an off-set of the linear extrapolation to $n=0$, but may be neglected if $v_3(T)$ is small.

In Fig.~\ref{fig:v2vsn05}, we show this generalized virial plot for $v_2^{\rm eff}(T,n)$. The dashed line corresponds to $v_2(T)+v_3(T) n_{\rm B}^{1/2} \ln (4 \pi n_{\rm B}/T_{\rm Ha}^2) + v_4(T) n_{\rm B}^{1/2}$ with the approximation (\ref{v4approx}) for $v_4(T)$. Because $v_3(100)$ is very small, the off-set at very low densities is not seen.
For comparison, the interpolation formula is also shown, and a linear behavior is seen. 
The extrapolation to $n=0$ misses the exact value $v_2(100)$ as also discussed above. 
There it was argued that the interpolation formula does not contain this benchmark by construction. 
However, if we assume that the interpolation formula gives a reasonable approximation in a wide range of parameter values, 
the linear behavior in the generalized virial plot is clearly seen. 
The extracted slope $v_4^{\rm GDSMFB}(100)= 0.00135$ is in good agreement with the value 0.00129 from the approximation (\ref{v4approx}). 

In addition to the isotherm $T_{\rm Ha}=100$, we studied also other isotherms ranging from $T_{\rm Ha}=50$ to $T_{\rm Ha} = 400$.
The extracted slope $v_4^{\rm GDSMFB}(T)$ show the $1/T^2$ behavior in accordance with Eq. (\ref{v4approx}).

\subsection{The $n^{5/2}$ term}

The investigation of the uniform electron gas is of interest not only for the discussion of the exchange-correlation term of the energy-density functional in DFT calculations, for which  analytical formulae have been derived by Groth, Dornheim, and Bonitz \cite{review,GDB2017}.
It is also a prerequisite for the treatment of the more interesting case of a two-component plasma, e.g., the hydrogen plasma.
For instance, the equation of state at low densities is of interest in helioseismology  \cite{Daeppen1988} where the fourth virial coefficient $v_4(T)$ is relevant \cite{DeWitt1998}.
In this context, the high-temperature limit of $v_2(T \to \infty)$ and the relation to $v_4(T)$ has been discussed in Refs.~\cite{KKR15,Dornheim_HEDP_2022}. 
For a discussion of the fourth virial coefficient $v_4(T)$ of the hydrogen plasma see also Alastuey and Ballenegger \cite{AB10,AB12}. 
The correct determination of the fourth virial coefficient $v_4(T)$ of the UEG is an important prerequisite for finding expressions for the fourth virial coefficient $F_4(T)$ in the free energy (\ref{Fvir}) associated with the density power $n^{5/2}$. However, we leave the discussion of this question to future work.

\section{Conclusions}\label{sec7}

Quantum statistics gives us exact expressions for  thermodynamic and transport properties of plasmas in terms of equilibrium correlation functions,
but their evaluation is a complex problem in many-particle physics. 
Numerical simulations are becoming more accurate as computer capacity increases.
However, they need to be checked for their limitations, such as size effects, but also for fundamental problems such as the correct description of electron-electron collisions in the framework of DFT or strategies to deal with the sign problem in PIMC simulations.
PIMC simulations are expected to provide an adequate description of electron-electron interactions, but are currently unable to solve complex plasmas such as multiply charged ions at low temperatures.

The use of analytical results for the virial expansion of thermodynamic properties as a benchmark for  PIMC calculations for the uniform electron gas is demonstrated.
In particular, we show that high-precision PIMC simulations confirm the correct form of the virial expansion that has been recently discussed \cite{Dornheim_HEDP_2022}. 
It also seems possible to obtain numerical values for higher virial coefficients, in particular the interesting virial coefficient $v_4(T)$ for the order $n^{5/2}$ of the 
free energy. 
These values can be considered as exact results in plasma physics. 

Analytical theory gives us exact results in limiting cases as benchmarks. 
These can be used to obtain results for parameter ranges where numerical simulations are not efficient, e.g. in the range of low densities. 
Virial expansions are used to control theories and numerical simulations. They are of interest for the construction of interpolation formulas.

The UEG is a comparatively simple case where PIMC simulations are possible with high accuracy. It will be interesting to extend the present considerations to a two-component system like the hydrogen plasma~\cite{Militzer_PRE_2001,Bohme_PRL_2022,filinov2023equation,Hamann_PRR_2023,filinov2023equation}, the positronium plasma or the electron-hole plasma.

\appendix

\section{Parameter values and units}
\label{App:1}
It is convenient to introduce dimensionless variables instead of $T,n$.
We use atomic units with the Hartree energy
\begin{equation}
 E_{\rm Ha}=\left(\frac{e^2}{4 \pi \epsilon_0}\right)^2 \frac{m}{\hbar^2}=27.21137\,{\rm eV} =2\, {\rm Ry},
\end{equation}
and the Bohr radius
\begin{equation}
 a_{\rm B}=\frac{4 \pi \epsilon_0}{e^2} \frac{\hbar^2}{m}=5.2918 \times 10^{-11}\,{\rm m}.
\end{equation}
The density in atomic units is usually represented by the radius of a sphere containing an electron,
\begin{equation}
 r_s= \left(\frac{3}{4 \pi n}\right)^{1/3}\frac{1}{a_{\rm B}}.
\end{equation}
The temperature is related to the energy $k_{\rm B}T$, so that  1~eV corresponds to 11604.6~K.
We denote $T_{\rm eV}$ as $k_{\rm B}T$ measured in units of eV, $T_{\rm Ha}$ in units of $E_{\rm Ha}$, and $T_{\rm Ry}$ in units of Ry so that
\begin{equation}
 T_{\rm Ha}=\frac{k_{\rm B}T}{E_{\rm Ha}}= 2 T_{\rm Ry}=27.21137\, T_{\rm eV}.
\end{equation}
Another well-known choice of dimensionless parameters is
\begin{equation}
 \Gamma = \frac{e^2}{4\pi\epsilon_0 k_{\rm B}T} \left(\frac{4\pi}{3} n \right)^{1/3},\qquad  \Theta=\frac{2mk_{\rm B}T}{\hbar^2}(3 \pi^2 n)^{-2/3}.
\end{equation}
The plasma parameter $\Gamma$ characterises the ratio of potential to kinetic energy in the non-degenerate case, 
and the electron degeneracy parameter $\Theta$ characterises the range in which the electrons are degenerate.
Different sets of dimensionless parameters are related. 
Thus, PIMC calculations are performed for specific parameter values of $r_s, \Theta$,
the corresponding plasma parameters $n,T$ are determined as follows,
\begin{equation}
 n= \frac{3}{4 \pi}\frac{1}{(r_sa_{\rm B})^3},\qquad k_{\rm B}T=E_{\rm Ha}\frac{1}{2}\left(\frac{9 \pi}{4}\right)^{2/3} \frac{\Theta}{r_s^2}
\end{equation}
with $E_{\rm Ha}/k_{\rm B}= 315777.1$ K.

\section{Parameter values for the GDSMFB interpolation formula (\ref{GDSMFB})}
\label{sec:Parameter}
The coefficients $a,b,c,d,e$ are again Pad\'e formulae with respect to temperature given in the Supplemental material to  \cite{groth_prl}. We give the expressions for the unpolarized case ($\xi = 0$),
\begin{eqnarray}
    a(\theta) &=& 0.610887 \tanh\left(\frac{1}{\theta}\right) \frac{a_1 + a_2\theta^2 - 0.09227\theta^3 + a_3\theta^4}{1 + a_4\theta^2 + a_5\theta^4}, \nonumber\\
    b(\theta) &=& \tanh\left(\frac{1}{\sqrt{\theta}}\right)\frac{b_1 + b_2\theta^2 + b_3\theta^4}{1 + b_4\theta^2 + b_5\theta^4}, \nonumber\\
    c(\theta) &=& (c_1 + c_2 e^{-1/\theta})e(\theta), \\
    d(\theta) &=& \tanh\left(\frac{1}{\sqrt{\theta}}\right)\frac{d_1 + d_2\theta^2 + d_3\theta^4}{1 + d_4\theta^2 + d_5\theta^4}, \nonumber\\
    e(\theta) &=& \tanh\left(\frac{1}{\theta}\right)\frac{e_1 + e_2\theta^2 + e_3\theta^4}{1 + e_4\theta^2 + e_5\theta^4}.\nonumber
    \label{eq:pade_coeff}
\end{eqnarray}
The paramters involved in those Pad\'e formulae are summarised below in Table \ref{tab:pade_parameter}.
\begin{table}
    \caption{Table for the parameters for the Pad\'e coefficients. e.g. the value in the first row and first column corresponds to $b_1$ and so on.}\label{tab:pade_parameter}
    \begin{center}
        \begin{tabular}{lS[table-format=1.5]S[table-format=2.6]S[table-format=-1.6]S[table-format=1.6]S[table-format=2.6]}
            \toprule
            \mc{Sub}& \mc{a}&\mc{b} & \mc{c} & \mc{d} & \mc{e} \\
            \hline
            1 & 0.75&0.343690 & 0.875944 & 0.727009 & 0.253882 \\
            2 & 3.04363& 7.821595 & -0.230131 & 2.382647 & 0.815795\\
            3 & 1.7035& 0.300484 &  & 0.302212 & 0.064684 \\
            4 & 8.31051& 15.844347 & & 4.393477 & 15.098462 \\
            5 & 5.1105 & 2.350479 & & 0.729951 & 0.230761 \\
            \toprule
        \end{tabular}
    \end{center}

\end{table}

\section{Fourth virial coefficient from interpolation formula}\label{app:3}

We give some values for the virial expansion (\ref{v0123}) and the effective virial coefficients (\ref{v2eff1}), (\ref{v3eff}), derived from the GDSMFB interpolation formula, in Tab. \ref{tab:GDSMFB}.

\label{App:v4}
\begin{table*}[htp]
\caption{Density-dependent GDSMFB data, corresponding parameter values and virial coefficients (atomic units), $T_{\rm Ha}=100$.}
\begin{center}
\begin{tabular}{S[table-format=2.1]S[table-format=4.4]cS[table-format=1.6]S[table-format=-1.6]S[table-format=1.6]S[table-format=-1.6]S[table-format=-1.6]S[table-format=1.6]S[table-format=-1.5e-1]}
\toprule
\mc{$r_s$}& \mc{$\Theta$} & \mc{$T_{\rm Ha}$} & \mc{$n_{\rm B}$} & \mc{$v^{\rm GDSMFB}$} & \mc{$n_{\rm B}^{1/2}$} &\mc{$-1/\ln(4 \pi n_{\rm B}/T_{\rm Ha}^2)$} & \mc{$v_2(T_{\rm Ha})$} & \mc{$v_2^{\rm eff, GDSMFB}$}& \mc{$v_3^{\rm eff,GDSMFB}$}  \\ 
\hline 
10 & 5430.11 & 100 & 0.000239 & -0.002739 & 0.015451 & 0.066580 & -0.01147 & -0.014702 	& 1.39248e-2\\ 
4 & 868.817 & 100 & 0.003730 & -0.010825 & 0.061075 &0.814955 & -0.01147 & -0.012858 	&1.85162e-3\\ 
2 & 217.204 & 100 & 0.029842 & -0.030937 & 0.172747 & 0.981242 &  -0.01147 & -0.012273 		&4.55733e-4\\ 
1.6 & 139.011 & 100 & 0.058284 & -0.042791 & 0.241421 & 0.105023 & -0.01147 & -0.012102	& 2.75842e-4\\ 
1 & 54.3011 & 100 & 0.238732 & -0.089082 & 0.488603 &0.123278  &  -0.01147 & -0.011658 	&4.73957e-5\\ 
0.8 & 34.7527 & 100 & 0.466274 & -0.12578 & 0.682843 &  0.134367&  -0.01147 & -0.011356 	&-2.26029e-5\\ 
0.5 & 13.5753 & 100 & 1.90986 & -0.263072 & 1.38198 & 0.165775 &  -0.01147 & -0.010437 	&-1.24015e-4\\ 
\toprule
\end{tabular} 
\end{center}
\label{tab:GDSMFB}
\end{table*}

The total contributions of different orders to the virial expansion of the potential energy $v$ are shown in Tab. \ref{tab:4}.
The high-temperature range is considered, $T_{\rm Ha}=100$.
The contributions of the odd orders virial terms containing $v_1, v_3$ at high temperatures are small compared with the even terms $v_0, v_2, v_4$.
\begin{table*}[htp]
\caption{Virial expansion: The total contribution of different orders (in [Ha]) to the virial expansion of the potential energy $v$, $T_{\rm Ha}=100$. }
\begin{center}
 \begin{tabular}{@{}S[table-format=2.1]cS[table-format=-1.6]S[table-format=-1.6]S[table-format=1.4e-1]S[table-format=-1.5e-1]S[table-format=1.5e-2]S[table-format=-1.5e-1]}
\toprule
\mc{$r_s$}  &  \mc{$T_{\rm Ha}$}& \mc{$v^{\rm GDSMFB}$ [Ha]}  & \mc{$v_0n_{\rm B}^{1/2}$}& \mc{$ v_1 n_{\rm B} \ln(4 \pi n_{\rm B}/T_{\rm Ha}^2)$} & \mc{$v_2 n_{\rm B}$} &\mc{$v_3 n_{\rm B}^{3/2}  \ln(4 \pi n_{\rm B}/T_{\rm Ha}^2)$}&\mc{$v_4^{\rm eff,GDSMFB}n_{\rm B}^{3/2}$}\\
\hline
10  &100      &-0.002741    &-0.002739  &5.6323e-7  &-2.73838e-6  &4.62741e-11       &-7.71503e-7   \\
4   &100    &-0.010866    &-0.010825  &7.1898e-6  &-4.27872e-5 &2.33496e-9 	& -5.17859e-6    \\
2   & 100  &-0.030937    &-0.030619  &4.7771e-5  &-3.42297e-4  &4.38806e-8    &-2.39862e-5    \\
1.6 & 100  & -0.043409 & -0.042791	& 8.7174e-5	& -6.6855e-4	& 1.11908e-7	& -3.69355e-5	\\
1  & 100 & -0.089082&	-0.086602	&3.0419e-4	&-2.73838e-3	&7.90308e-7	&4.56357e-5\\ 
0.8  & 100  & -0.12578 & -0.121031	&5.4509e-4	&-5.3484e-3	&1.97918e-6	&5.15803e-5\\
0.5  & 100  & -0.263072 & -0.244949	&1.8097e-3	&-2.1907e-2	&1.32984e-5	&1.96121e-3 \\
\toprule
\end{tabular}
\end{center}
\label{tab:4}
\end{table*}

\section*{Acknowledgments}
This work was partially supported by the Center for Advanced Systems Understanding (CASUS) which is financed by Germany’s Federal Ministry of Education and Research (BMBF) and by the Saxon state government out of the State budget approved by the Saxon State Parliament. 
This work has received funding from the European Research Council (ERC) under the European Union’s Horizon 2022 research and innovation programme
(Grant agreement No. 101076233, "PREXTREME").
This work was partially performed on the HoreKa supercomputer funded by the
Ministry of Science, Research and the Arts Baden-Württemberg and by
the Federal Ministry of Education and Research, and at the Norddeutscher Verbund f\"ur Hoch- und H\"ochstleistungsrechnen (HLRN) under grant mvp00024. G.R. acknowledges a fellowship of the Alexander von Humboldt programme of the Foundation for Polish Science.


\bibliography{bibliography}

\end{document}